\documentclass[runningheads]{llncs}
\usepackage[T1]{fontenc}
\usepackage{graphicx}
\usepackage{amsmath}
\usepackage{amssymb}
\usepackage{amsfonts}
\usepackage{booktabs}
\usepackage[table,xcdraw]{xcolor}
\usepackage{multirow}
\usepackage[colorlinks=true,allcolors=blue]{hyperref}
\usepackage{bbding} 
\begin{document}
\title{Towards Arbitrary-scale Histopathology Image Super-resolution: An Efficient Dual-branch Framework based on Implicit Self-texture Enhancement}
\titlerunning{ISTE}

\author{Linhao Qu$^*$\inst{1}, Minghong Duan$^*$\inst{1}, Zhiwei Yang\inst{1,2}, Manning Wang\textsuperscript{\Envelope}\inst{1}, Zhijian Song\textsuperscript{\Envelope}\inst{1}}
\authorrunning{Qu et al.}

\institute{Digital Medical Research Center, School of Basic Medical Science, Fudan University,
Shanghai Key Lab of Medical Image Computing and Computer Assisted Intervention, Shanghai 200032, China,
 \{mnwang, zjsong\}@fudan.edu.cn
 \and Academy for Engineering \& Technology, Fudan University, Shanghai 200433, China
 }

\renewcommand{\thefootnote}{}
\footnotetext{$^*$Linhao Qu and Minghong Duan contributed equally to this work.}
\maketitle              
\begin{abstract}
Existing super-resolution models for pathology images can only work in fixed integer magnifications and have limited performance. 
Though implicit neural network-based methods have shown promising results in arbitrary-scale super-resolution of natural images, it is not effective to directly apply them in 
pathology images, because pathology images have special fine-grained image textures different from natural images. To address this challenge, we propose a dual-branch framework 
with an efficient self-texture enhancement mechanism for arbitrary-scale super-resolution of pathology images. Extensive experiments on two public datasets show that our method 
outperforms both existing fixed-scale and arbitrary-scale algorithms. To the best of our knowledge, this is the first work to achieve arbitrary-scale super-resolution in the field 
of pathology images. Codes will be available.

\end{abstract}
\section{Introduction}
High-resolution pathology Whole Slide Images (WSIs) contain rich cellular morphology and pathological patterns, which are the basis for a series of automated pathology image analysis 
tasks \cite{1,2,3,4}. However, the acquisition and use of high-quality WSIs remain limited in the daily clinical workflow \cite{18}. On one hand, high-resolution WSIs need to be acquired by high 
magnification scanners \cite{4}, which are expensive and time-consuming to use. On the other hand, high-resolution WSIs are very large, which increases the difficulty of data storage and 
management \cite{6,8,32}. Therefore, generating high-resolution images from low-resolution ones using software algorithms will largely facilitate the practical clinical analysis of pathological images \cite{6,18}.

\begin{figure*}[t]
    \centering
    \includegraphics[scale=0.17]{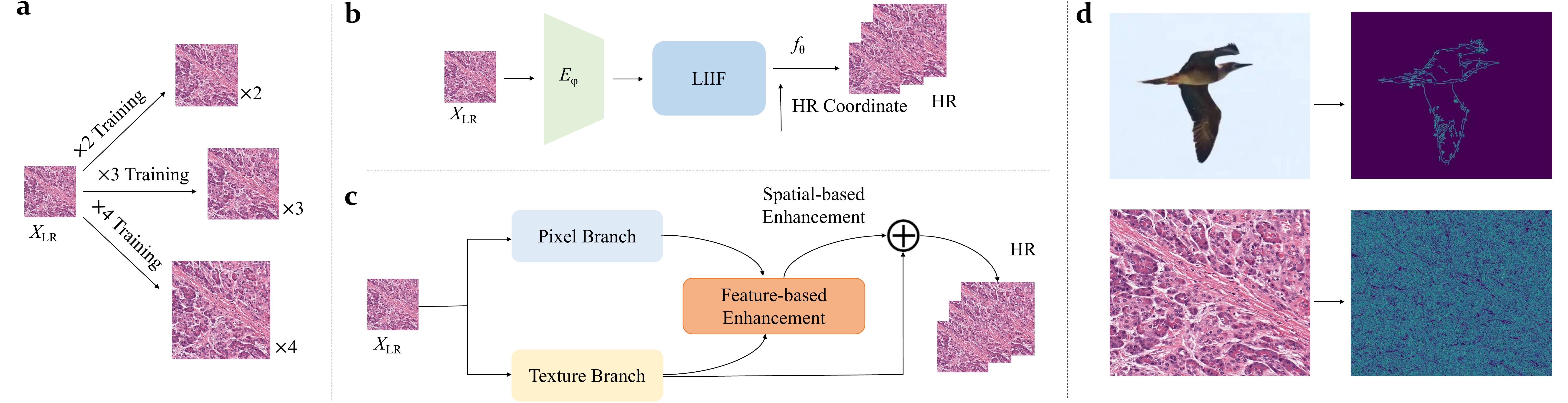}
    \caption{Motivation and architecture of the proposed ISTE. (a) All existing studies for pathology image super-resolution can only be trained and work at fixed integer scaling factors.
    (b) Some natural image super-resolution algorithms based on implicit neural networks \cite{20} can perform arbitrary-scale super-resolution, 
    but cannot properly handle WSI textures. (c) The overall architecture of ISTE, which consists of two branches and two specially-designed texture enhancement strategies. 
    (d) Natural and pathology images and their textures extracted by using canny operator. It can be seen that, in contrast to natural images, pathology images contain richer 
    textures of fine-grained cell morphology with special spatial arrangement.}
    \label{figure1}
\end{figure*}

Super-resolution (SR) technique is an effective way to solve this problem \cite{11,12,13,14,15,30,34,36,37}. In the field of pathological image analysis, several studies \cite{16,17,18} apply convolutional 
neural networks to perform SR. As shown in Fig. \ref{figure1} (a), although these methods achieve good performance, they can only be trained and tested at a specific integer scale. However, pathologists usually 
need to continuously zoom in and out at different magnifications to perform diagnosis, so an arbitrary-scale super-resolution method is preferable. Unfortunately, to the best of our knowledge, 
there are currently no models that can achieve arbitrary-scale SR in the pathology field. Recently, inspired by implicit neural networks \cite{19}, some studies \cite{20,21} have pioneered the arbitrary-scale SR 
for natural images. As shown in Fig. \ref{figure1} (b), although these studies can be directly applied to pathology image SR, they are not effective in handling the special textures in WSIs, which is crucial in 
SR for pathology images. As shown in Fig. \ref{figure1} (d), unlike natural images, pathology images contain richer fine-grained cell morphology textures with special spatial arrangements.

To address this challenge, an efficient dual-branch framework based on \textbf{I}mplicit \textbf{S}elf-\textbf{T}exture \textbf{E}nhancement (called ISTE) is proposed in this paper for arbitrary-scale SR of pathology images. 
Fig. \ref{figure1} (c) briefly illustrates the overall architecture of ISTE. Specifically, ISTE contains a pixel learning branch and a texture learning branch, both of which are based on 
implicit neural networks \cite{20}, thus enabling the magnification of images of arbitrary scales. In the pixel learning branch, we propose the Local Feature Interactor module to obtain 
richer pixel features and in the texture learning branch, we propose the Texture Learner module to enhance the network's learning of texture information. After that, we design two 
texture enhancement strategies, namely the feature-based texture enhancement and the spatial domain-based texture enhancement, to further improve the textures of the output images. 
Extensive experiments on two public datasets have shown that ISTE provides better performance than existing fixed-magnification and arbitrary-magnification algorithms at multiple scales. 
To the best of our knowledge, this is the first work to achieve arbitrary-scale super-resolution in the field of pathology images.

\section{Method}
\subsection{Framework Overview}
\begin{figure*}[t]
    \centering
    \includegraphics[scale=0.15]{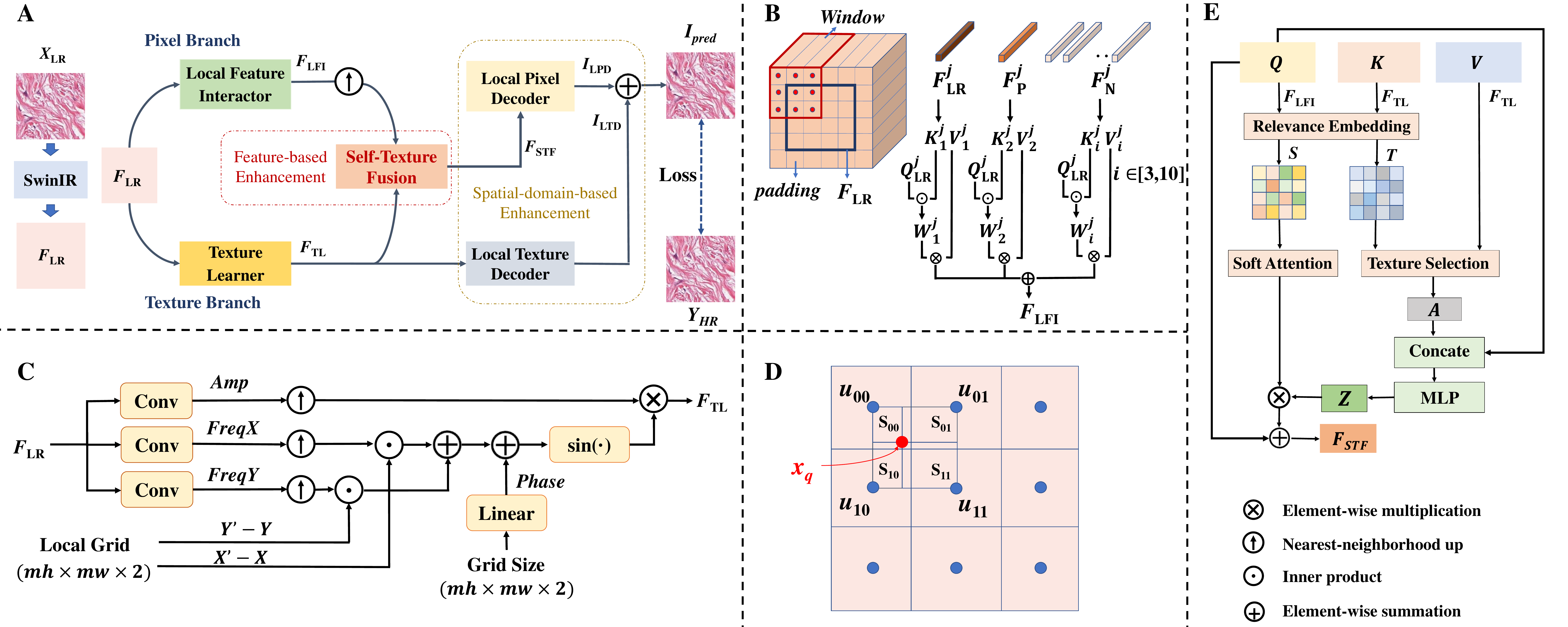}
    \caption{Detailed architecture of our ISTE (A) and its key components. B Local Feature Interactor, C Texture Learner, D Coordinate Diagram of $F_{\mathrm{STF}}$ and $F_{\mathrm{TL}}$ for Local Pixel Decoder and Local Texture Decoder. E Self-Texture Fusion Module.}
    \label{figure2}
\end{figure*}

Fig. \ref{figure2} A shows the detailed architecture of our proposed ISTE. SwinIR \cite{22,23} without upsampling layers is first used to extract features from the input low-resolution image $X_{\mathrm{LR}}$ and 
then the extracted feature map $F_{\mathrm{LR}}$ is input into the pixel branch and the texture branch, respectively. Both branches are based on implicit neural networks \cite{20}, thus enabling the 
magnification of arbitrary-scale images. In the pixel learning branch, the Local Feature Interactor (LFI) module is used to enhance the network's perception and interaction of local pixel 
features and obtain a richer pixel feature $F_{\mathrm{LFI}}$. In the texture learning branch, the Texture Learner (TL) module is used to enhance the network's learning of texture information and extract 
the texture feature $F_{\mathrm{TL}}$. After that, we utilize a two-stage texture enhancement strategy to process the output features from the two branches, where the first stage is feature-based texture 
enhancement and the second stage is spatial domain-based texture enhancement. Considering that pathology images contain many similar cell morphology and periodic texture patterns, we design 
the Self-Texture Fusion (STF) module and obtain $F_{\mathrm{STF}}$ to accomplish feature-based texture enhancement. After that, we obtain the output high-resolution image through spatial-domain-based texture 
enhancement. Specifically, we decode the enhanced feature $F_{\mathrm{STF}}$ through the Local Pixel Decoder to obtain the spatial image $I_{\mathrm{LPD}}$. At the same time, we use Local Texture Decoder to decode the 
texture feature $F_{\mathrm{TL}}$ to spatial texture $I_{\mathrm{LTD}}$. Finally, $I_{\mathrm{LTD}}$ and $I_{\mathrm{LPD}}$ are added up through element-wise summation to obtain the output high-resolution image $I_{\mathrm{pred}}$. Then, we use L1 loss to calculate 
the loss between it and the ground truth image. The LFI, TL and STF modules in our ISTE framework are presented in Sections 2.2, 2.3 and 2.4, respectively. The Local Pixel Decoder and Local Texture Decoder are introduced in Section 2.5.

\subsection{Local Feature Interactor}
As shown in Fig. \ref{figure2} B, the LFI module first assigns a window of size 3$\times$3 to each vector of the feature map $F_{\mathrm{LR}}$ of size $h\times w\times64$, and we denote each vector of 
$F_{\mathrm{LR}}$ as $F_{LR}^j\ (j=1,2,...,h\times w)$. Eight neighboring vectors in the window around $F_{LR}^j$ form a set $F_N^j=\{F_{N_i}^j|\ i=3,4,...,10\}$, and the pooling vector of the window is denoted 
as $F_P^j$. The feature map $F_{\mathrm{LFI}}$ output by the LFI module is calculated through self-attention so that each point on the feature map incorporates local features while paying more attention to 
itself. We denote each vector of $F_{\mathrm{LFI}}$ as $F_{LFI}^j\ (j=1,2,...,h\times w)$, and it is calculated through Equation \ref{eq1}.
\begin{equation}
    F_{L F I}^j=\sum_{i=1}^{10} \frac{\exp \left(\left(Q_{L R}^j\right)^T K_i^j\right)}{\sqrt{d} \sum_{i=1}^{10} \exp \left(\left(Q_{L R}^j\right)^T K_i^j\right)} V_i^j,  \label{eq1}
\end{equation}
where $Q_{LR}^j$ is the Query mapped linearly from $F_{LR}^j$, $K_1^j$ is the Key mapped linearly from $F_{LR}^j$, $V_1^j$ is the Value mapped linearly from $F_{LR}^j$, $K_2^j$ is the Key mapped 
linearly from $F_P^j$, $V_2^j$\ is the Value mapped linearly from $F_P^j$, $\{K_i^j|\ i=3,4,...,10\}$ is the Key mapped linearly from $F_N^j$, $\{V_i^j|\ i=3,4,...,10\}$ is the Value mapped linearly 
from $F_N^j$, and $d$ is the dimension of these vectors. The parameters used by each window are shared in Self Attention calculation.

\subsection{Texture Learner}
Inspired by LTE \cite{21}, Texture Learner (TL) is proposed for learning texture information in pathology images. We use $sine$ activation to help our model represent the cell and texture that appear periodically in pathology images. Specifically, we normalize the value of 2D 
coordinate $(X',\ Y')=\{({x'}_i,{y'}_j)|i=1,2,...,mw,j=1,2,...,mh\}$ in the continuous HR image domain and the value of 2D coordinate $(X,\ Y)=\{(x_i,y_j)|i=1,2,...,mw,j=1,2,...,mh\}$ 
nearest to $(X',\ Y')$ in the continuous LR image domain between -1 and 1, and the Local Grid is defined as $(X'-X,\ Y'-Y)$. Since each coordinate of the HR image has a 
corresponding coordinate in the LR image that is closest to it, the number of both the HR and LR image coordinate is equal to $mh \times mw$, where $m$ represents magnification. 
As shown in Fig. \ref{figure2} C, the TL outputs three feature maps of size $h \times w \times 256$ through three 3 $\times$ 3 convolutional kernels, and predicts the feature vectors $Amp$, $FreqX$, and $FreqY$ corresponding to each coordinate of the HR image through nearest neighbor interpolation. 
We use an MLP and Sigmoid activation function to map $\left(2/mw,2/mh\right)$ to a 256-dimensional feature vector $Phase$ to simulate the effect of texture fragment offset when the image scaling factor changes. The output of the FL module is calculated by Equation \ref{eq2}.
\begin{equation}
    F_{T L}=Amp \otimes Sin\left(FreqX \odot\left(X^{'}-X\right)+FreqY \odot\left(Y^{'}-Y\right)+Phase\right)  \label{eq2}
\end{equation}
where $\otimes$\ denotes the element-wise multiplication and $\odot$\ denotes the inner product.

\subsection{Self-Texture Fusion Module}

Inspired by STSRNet \cite{26} and T2Net \cite{27}, we propose a cross attention-based Self-Texture Fusion (STF) module. As shown in Fig. \ref{figure2} E, we use the features sampled by nearest-neighborhood interpolation 
from $F_{\mathrm{LFI}}$ as the Query (Q) of the cross attention module and the $F_{\mathrm{TL}}$ as the Key (K) and Value (V) of the cross attention module. To retrieve the texture features that are most relevant to the 
pixel feature $F_{\mathrm{LFI}}$ from $F_{\mathrm{TL}}$, we first calculate the similarity matrix $R$ of $Q$ and $K$, where each element $r_{i,j}$ of $R$ is calculated as Equation \ref{eq3}, where $q_i$ is an element of $Q$ and $k_j$ is an element of $K$. Then we 
obtain the coordinate index matrix $T$ with the highest similarity to $q_i$ in $K$. An element in $T$ is $t_i=\arg \max _j\left(r_{i, j}\right)$, and $t_i$ represents the position coordinates of the 
texture feature $k_j$ with the highest similarity to $q_i$ in $F_{\mathrm{TL}}$. We pick the feature vector $a_i$ with the highest similarity to each element in $Q$ from $V$ according to the coordinate index matrix $T$ to 
obtain the retrieved texture feature $A$, which can be represented by $a_i=v_{t_i}$, where $a_i$ is an element in $A$ and $v_{t_i}$ represents the element at the $t_i$-th position in $V$. In order to fuse the 
retrieved texture feature $A$ with the pixel feature $Q$, we first concatenate $Q$ with $A$ and obtain the aggregated feature $Z$ through the output of an MLP, that is $Z=M L P(\text { Concat }(Q, A))$. 
Finally, we calculate the soft attention map $S$, where an element $s_i$ in $S$ represents the confidence of each element $a_i$ in the retrieved texture 
feature $A$, and $s_i=\max _j\left(r_{i, j}\right)$. $F_{\mathrm{STF}}$ is calculated as Equation \ref{eq4}.

\begin{equation}
    r_{i, j}=\left\langle\frac{q_i}{\left\|q_i\right\|}, \frac{k_j}{\left\|k_j\right\|}\right\rangle,  \label{eq3}
\end{equation}

\begin{equation}
    F_{S T F}=Q \oplus Z \otimes S  \label{eq4}
\end{equation}
where $\langle\cdot\rangle$ denotes the inner product, $\|\cdot\|$ denotes the L2 norm, and $\oplus$ denotes the element-wise summation.

\subsection{Local Pixel Decoder and Local Texture Decoder}
Inspired by LIIF \cite{20}, we propose the Local Pixel Decoder (LPD) module to decode the feature $F_{\mathrm{STF}}$, into the pixel RGB value $I_{\mathrm{LPD}}$. 
We parameterize LPD as an MLP $f_\theta$, where $\theta$ is the network parameter. As shown in Fig. \ref{figure2} (D), $u$ denotes the coordinates of $F_{\mathrm{LR}}$ and $x_q$ denotes the coordinate of $F_{\mathrm{STF}}$ and $F_{\mathrm{TL}}$. 
We use $u_t\ (t\in{00,01,10,11})$ to denote the upper-left, upper-right, lower-left, and lower-right coordinates of an arbitrary point $x_q$, respectively, and the RGB value of $x_q$ in the HR image 
LPD can be expressed by Equation \ref{eq5}, where $I_{LPD}(x_q)$ is the RGB value of $x_q$ in image $I_{\mathrm{LPD}}$, $c$ contains two elements, $2/mh$ and $2/mw$, which are the pixel size of $I_{\mathrm{LPD}}$ and facilitate the 
decoding of pixel information. Similarly, we parameterize the $I_{\mathrm{LTD}}$ as an MLP $g_\varphi$, and the RGB values corresponding to the $I_{\mathrm{LTD}}$ coordinates $x_q$ of the texture-informed image output by the 
spatial domain-based texture enhancement can be caluclated by Equation \ref{eq6}. We decode the texture features into texture information $I_{\mathrm{LTD}}$ by the LTD module and add it to $I_{\mathrm{LPD}}$ to achieve the 
texture enhancement, where $\varphi$ is the network parameter of the MLP $g_\varphi$, $S_t\ (t\in{00,01,10,11})$ is the area of the rectangular region between $x_q$ and $u_t$, and the weights are 
normalized by $S=\sum_{t \in\{00,01,10,11\}} S_t$.
\begin{equation}
    I_{L P D}\left(x_q\right)=\sum_{t \in\{00,01,10,11\}} \frac{s_t}{s} \cdot f_\theta\left(F_{S T F}, x_q-u_t, c\right),  \label{eq5}
\end{equation}

\begin{equation}
    I_{L T D}\left(x_q\right)=\sum_{t \in\{00,01,10,11\}} \frac{s_t}{s} \cdot g_{\varphi}\left(F_{T L}\right),  \label{eq6}
\end{equation}

\section{Experiments}
\subsection{Datasets, Competitors, Metrics and Implementation Details}
We used TMA Dataset \cite{28,18} and TCGA lung cancer dataset for experiments. For the TMA Dataset, we randomly selected 460 WSIs (average 3850$\times$3850 pixels each) as the training set, 57 WSIs as the validation set, and 56 WSIs as the test set. 
For the TCGA dataset, we selected 5 slides (average 100000$\times$100000 pixels each) cut into 400 sub-images of 3072$\times$3072, and randomly selected 320 as the training set, 40 as the validation set, and 40 as the test set.

We compared the performance of ISTE with SOTA SR methods in both the pathology image domain: Li et al. \cite{18}, and the natural image domain: Bicubic, EDSR \cite{29}, LIIF \cite{20} and LTE \cite{21}, where 
the latter two are implicit neural network-based methods. For a fair comparison, the backbone used for LIIF and LTE is also SwinIR \cite{23} without upsampling layers. Following these studies, 
our evaluation metrics include structural similarity (SSIM), peak signal-to-noise ratio (PSNR), and Frechet Inception Distance (FID).

Following LIIF \cite{20}, we used a 48$\times$48 patch as the input for training. We first randomly sampled the magnification $m$ in a uniform distribution $U (1, 4)$, and 
cropped patches with size of 48$m$ $\times$ 48$m$ from training images in a batch. We then resized the patches to 48 $\times$ 48 and did a Gaussian blur to simulate degradation. 
For the ground truth images, we sampled ${48}^2$ pixels from the corresponding cropped patches to form RGB-Coordinate pairs. 
We used PyTorch with Adam as the optimizer, setting the initial learning rate to 0.0001 and epochs to 1000.

\subsection{Experimental Results}
We compared our ISTE with competitors at five magnifications of $\times$2, $\times$3, $\times$4, $\times$6, and $\times$8. Table \ref{table1} show the results on the TMA and TCGA datasets. 
As can be seen, ISTE achieves the highest performance in almost all metrics at all magnifications. It is worth noting that the existing SOTA methods in the field of 
pathology image SR, Li's method \cite{18} cannot achieve arbitrary-scale magnification, so they need to be retrained at each magnification, whereas our method needs only a 
one-time training. Even so, our method outperforms Li's method \cite{18} with a large margin. On the other hand, compared to the other two implicit neural network-based 
methods LIIF \cite{20} and LTE \cite{21} in the field of natural images, our method also shows superior performance in most of the metrics.

\begin{table}[t]
    \centering
    \caption{Performance comparison of the methods on the TMA and TCGA datasets.}
      \scalebox{0.59}{
\begin{tabular}{
>{\columncolor[HTML]{FFFFFF}}c |
>{\columncolor[HTML]{FFFFFF}}c |
>{\columncolor[HTML]{FFFFFF}}c 
>{\columncolor[HTML]{FFFFFF}}c 
>{\columncolor[HTML]{FFFFFF}}c 
>{\columncolor[HTML]{FFFFFF}}c 
>{\columncolor[HTML]{FFFFFF}}c 
>{\columncolor[HTML]{FFFFFF}}c |
>{\columncolor[HTML]{FFFFFF}}c 
>{\columncolor[HTML]{FFFFFF}}c 
>{\columncolor[HTML]{FFFFFF}}c 
>{\columncolor[HTML]{FFFFFF}}c 
>{\columncolor[HTML]{FFFFFF}}c 
>{\columncolor[HTML]{FFFFFF}}c 
>{\columncolor[HTML]{FFFFFF}}c 
>{\columncolor[HTML]{FFFFFF}}c 
>{\columncolor[HTML]{FFFFFF}}c }
\toprule    
\cellcolor[HTML]{FFFFFF}                          & \cellcolor[HTML]{FFFFFF}                         & \multicolumn{6}{c|}{\cellcolor[HTML]{FFFFFF}In-distribution}                                                                                                                                                                                                                               & \multicolumn{9}{c}{\cellcolor[HTML]{FFFFFF}Out-of-distribution}                                                                                                                                                                                                                                                                                                                                                                                                     \\ \cline{3-17} 
\cellcolor[HTML]{FFFFFF}                          & \cellcolor[HTML]{FFFFFF}                         & \multicolumn{3}{c|}{\cellcolor[HTML]{FFFFFF}×2}                                                                                                           & \multicolumn{3}{c|}{\cellcolor[HTML]{FFFFFF}×3}                                                                       & \multicolumn{3}{c|}{\cellcolor[HTML]{FFFFFF}×4}                                                                                                                     & \multicolumn{3}{c|}{\cellcolor[HTML]{FFFFFF}×6}                                                                                                                     & \multicolumn{3}{c}{\cellcolor[HTML]{FFFFFF}×8}                                                                          \\ \cline{3-17} 
\multirow{-3}{*}{\cellcolor[HTML]{FFFFFF}Dataset} & \multirow{-3}{*}{\cellcolor[HTML]{FFFFFF}Method} & PSNR ↑                                & FID↓                                 & \multicolumn{1}{c|}{\cellcolor[HTML]{FFFFFF}SSIM↑}                                  & PSNR↑                                 & FID↓                                 & SSIM↑                                  & PSNR↑                                 & FID↓                                  & \multicolumn{1}{c|}{\cellcolor[HTML]{FFFFFF}SSIM↑}                                  & PSNR↑                                 & FID↓                                  & \multicolumn{1}{c|}{\cellcolor[HTML]{FFFFFF}SSIM↑}                                  & PSNR↑                                 & FID↓                                   & SSIM↑                                  \\ \toprule
\cellcolor[HTML]{FFFFFF}                          & Bicubic                                          & 27.02                                 & 12.19                                & \multicolumn{1}{c|}{\cellcolor[HTML]{FFFFFF}0.8559}                                 & 24.17                                 & 39.41                                & 0.7289                                 & 22.62                                 & 69.32                                 & \multicolumn{1}{c|}{\cellcolor[HTML]{FFFFFF}0.6387}                                 & 20.94                                 & 117.22                                & \multicolumn{1}{c|}{\cellcolor[HTML]{FFFFFF}0.5426}                                 & 19.98                                 & 155.44                                 & 0.4972                                 \\
\cellcolor[HTML]{FFFFFF}                          & EDSR (CVPR2017)                                  & 28.79                                 & 6.97                                 & \multicolumn{1}{c|}{\cellcolor[HTML]{FFFFFF}0.8923}                                 & 23.60                                 & 24.33                                & 0.7259                                 & 23.88                                 & 49.68                                 & \multicolumn{1}{c|}{\cellcolor[HTML]{FFFFFF}0.7002}                                 & 21.79                                 & 88.92                                 & \multicolumn{1}{c|}{\cellcolor[HTML]{FFFFFF}0.5909}                                 & 20.65                                 & 114.05                                 & 0.5371                                 \\
\cellcolor[HTML]{FFFFFF}                          & LIIF (CVPR2021)                                  & 31.10                                 & 3.39                                 & \multicolumn{1}{c|}{\cellcolor[HTML]{FFFFFF}0.9427}                                 & 27.95                                 & 5.45                                 & 0.8764                                 & 25.99                                 & 15.71                                 & \multicolumn{1}{c|}{\cellcolor[HTML]{FFFFFF}0.8035}                                 & 23.64                                 & 50.76                                 & \multicolumn{1}{c|}{\cellcolor[HTML]{FFFFFF}{\color[HTML]{FF0000} \textbf{0.6833}}} & 22.20                                 & 80.61                                  & {\color[HTML]{FF0000} \textbf{0.6037}} \\
\cellcolor[HTML]{FFFFFF}                          & LTE (CVPR2022)                                   & 31.26                                 & 3.11                                 & \multicolumn{1}{c|}{\cellcolor[HTML]{FFFFFF}0.9432}                                 & 28.22                                 & 5.27                                 & 0.8787                                 & 26.25                                 & 14.37                                 & \multicolumn{1}{c|}{\cellcolor[HTML]{FFFFFF}0.8093}                                 & 23.75                                 & 51.78                                 & \multicolumn{1}{c|}{\cellcolor[HTML]{FFFFFF}0.6823}                                 & 22.18                                 & 80.98                                  & 0.5982                                 \\
\cellcolor[HTML]{FFFFFF}                          & Li's (MIA2021)                                   & 29.70                                 & 6.52                                 & \multicolumn{1}{c|}{\cellcolor[HTML]{FFFFFF}0.9106}                                 & 26.19                                 & 18.89                                & 0.8357                                 & 24.15                                 & 46.23                                 & \multicolumn{1}{c|}{\cellcolor[HTML]{FFFFFF}0.7537}                                 & 20.45                                 & 95.04                                 & \multicolumn{1}{c|}{\cellcolor[HTML]{FFFFFF}0.6188}                                 & 18.71                                 & 137.52                                 & 0.5425                                 \\ \cline{2-17} 
\multirow{-6}{*}{\cellcolor[HTML]{FFFFFF}TMA}     & \textbf{Ours}                                    & {\color[HTML]{FF0000} \textbf{32.71}} & {\color[HTML]{FF0000} \textbf{2.87}} & \multicolumn{1}{c|}{\cellcolor[HTML]{FFFFFF}{\color[HTML]{FF0000} \textbf{0.9445}}} & {\color[HTML]{FF0000} \textbf{28.77}} & {\color[HTML]{FF0000} \textbf{4.85}} & {\color[HTML]{FF0000} \textbf{0.8815}} & {\color[HTML]{FF0000} \textbf{26.70}} & {\color[HTML]{FF0000} \textbf{13.76}} & \multicolumn{1}{c|}{\cellcolor[HTML]{FFFFFF}{\color[HTML]{FF0000} \textbf{0.8142}}} & {\color[HTML]{FF0000} \textbf{23.90}} & {\color[HTML]{FF0000} \textbf{50.76}} & \multicolumn{1}{c|}{\cellcolor[HTML]{FFFFFF}0.6822}                                 & {\color[HTML]{FF0000} \textbf{22.22}} & {\color[HTML]{FF0000} \textbf{78.78}}  & 0.5953                                 \\ \toprule
\cellcolor[HTML]{FFFFFF}                          & Bicubic                                          & 28.89                                 & 15.94                                & \multicolumn{1}{c|}{\cellcolor[HTML]{FFFFFF}0.8791}                                 & 25.78                                 & 57.40                                & 0.7348                                 & 24.00                                 & 105.76                                & \multicolumn{1}{c|}{\cellcolor[HTML]{FFFFFF}0.6230}                                 & 22.02                                 & 190.59                                & \multicolumn{1}{c|}{\cellcolor[HTML]{FFFFFF}0.4959}                                 & 20.85                                 & 257.01                                 & 0.4347                                 \\
\cellcolor[HTML]{FFFFFF}                          & EDSR (CVPR2017)                                  & 30.80                                 & 5.59                                 & \multicolumn{1}{c|}{\cellcolor[HTML]{FFFFFF}0.9047}                                 & 27.37                                 & 46.01                                & 0.7803                                 & 25.43                                 & 112.04                                & \multicolumn{1}{c|}{\cellcolor[HTML]{FFFFFF}0.6776}                                 & 23.23                                 & 204.64                                & \multicolumn{1}{c|}{\cellcolor[HTML]{FFFFFF}0.5506}                                 & 21.90                                 & 241.72                                 & 0.4814                                 \\
\cellcolor[HTML]{FFFFFF}                          & LIIF (CVPR2021)                                  & 36.25                                 & 1.53                                 & \multicolumn{1}{c|}{\cellcolor[HTML]{FFFFFF}0.9716}                                 & 31.62                                 & {\color[HTML]{FF0000} \textbf{2.81}} & 0.9073                                 & 28.69                                 & 19.90                                 & \multicolumn{1}{c|}{\cellcolor[HTML]{FFFFFF}0.8187}                                 & 25.31                                 & 96.44                                 & \multicolumn{1}{c|}{\cellcolor[HTML]{FFFFFF}0.6564}                                 & 23.55                                 & 136.12                                 & 0.5545                                 \\
\cellcolor[HTML]{FFFFFF}                          & LTE (CVPR2022)                                   & 36.65                                 & 1.31                                 & \multicolumn{1}{c|}{\cellcolor[HTML]{FFFFFF}0.9729}                                 & 31.87                                 & 3.12                                 & 0.9094                                 & 28.95                                 & 18.61                                 & \multicolumn{1}{c|}{\cellcolor[HTML]{FFFFFF}0.8244}                                 & 25.44                                 & 92.17                                 & \multicolumn{1}{c|}{\cellcolor[HTML]{FFFFFF}0.6597}                                 & 23.62                                 & 125.45                                 & 0.5559                                 \\
\cellcolor[HTML]{FFFFFF}                          & Li's (MIA2021)                                   & 33.53                                 & 3.86                                 & \multicolumn{1}{c|}{\cellcolor[HTML]{FFFFFF}0.9472}                                 & 29.74                                 & 28.56                                & 0.8716                                 & 27.42                                 & 57.43                                 & \multicolumn{1}{c|}{\cellcolor[HTML]{FFFFFF}0.7844}                                 & 24.29                                 & 116.18                                & \multicolumn{1}{c|}{\cellcolor[HTML]{FFFFFF}0.6230}                                 & 22.83                                 & 140.12                                 & 0.5334                                 \\ \cline{2-17} 
\multirow{-6}{*}{\cellcolor[HTML]{FFFFFF}TCGA}    & \textbf{Ours}                                    & {\color[HTML]{FF0000} \textbf{38.28}} & {\color[HTML]{FF0000} \textbf{1.22}} & \multicolumn{1}{c|}{\cellcolor[HTML]{FFFFFF}{\color[HTML]{FF0000} \textbf{0.9748}}} & {\color[HTML]{FF0000} \textbf{32.46}} & 2.92                                 & {\color[HTML]{FF0000} \textbf{0.9111}} & {\color[HTML]{FF0000} \textbf{29.33}} & {\color[HTML]{FF0000} \textbf{17.42}} & \multicolumn{1}{c|}{\cellcolor[HTML]{FFFFFF}{\color[HTML]{FF0000} \textbf{0.8287}}} & {\color[HTML]{FF0000} \textbf{25.64}} & {\color[HTML]{FF0000} \textbf{88.00}} & \multicolumn{1}{c|}{\cellcolor[HTML]{FFFFFF}{\color[HTML]{FF0000} \textbf{0.6617}}} & {\color[HTML]{FF0000} \textbf{23.74}} & {\color[HTML]{FF0000} \textbf{119.98}} & {\color[HTML]{FF0000} \textbf{0.5578}} \\ \bottomrule
\end{tabular}}
    \label{table1}
\end{table}

\begin{figure*}[htbp]
    \centering
    \includegraphics[scale=0.25]{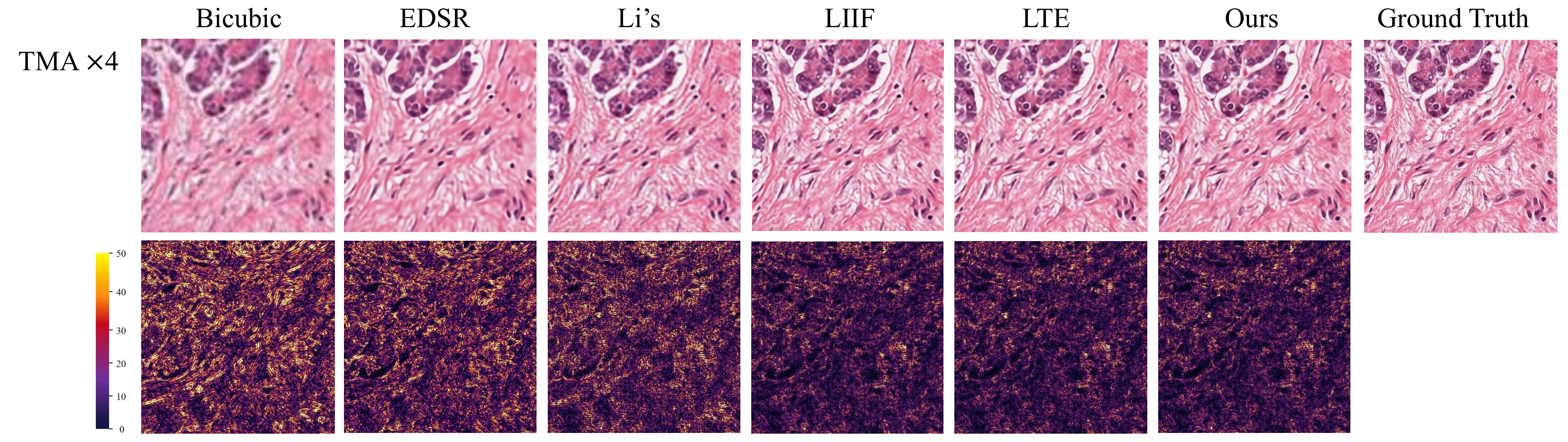}
    \caption{Visual comparison with error maps of different methods on the two datasets. The error maps represent the absolute 
    values of the error between the outputs and the ground truth images, with brighter colors representing larger errors.}
    \label{figure3}
\end{figure*}

Fig. \ref{figure3} shows the visualization of the super-resolution results and the error maps at $\times$4 magnification. It can be seen that ISTE recovers textures better and ISTE's results are more similar to 
the ground truth images. Fig. \ref{figure4} A shows the non-integeral magnification of LIIF and our ISTE. It can be seen that ISTE achieves arbitrary magnification with clear cell structure and 
texture features. In larger magnification, ISTE outperforms LIIF in terms of visual effect.
More visual comparisons with error maps of different methods on the TMA and TCGA datasets are shown in Fig. \ref{figure5}. 
More visual results of our method at arbitary magnifications on the TMA and TCGA datasets are shown in Fig. \ref{figure6}.

\section{Ablation Study}
\begin{figure*}[htbp]
    \centering
    \includegraphics[scale=0.14]{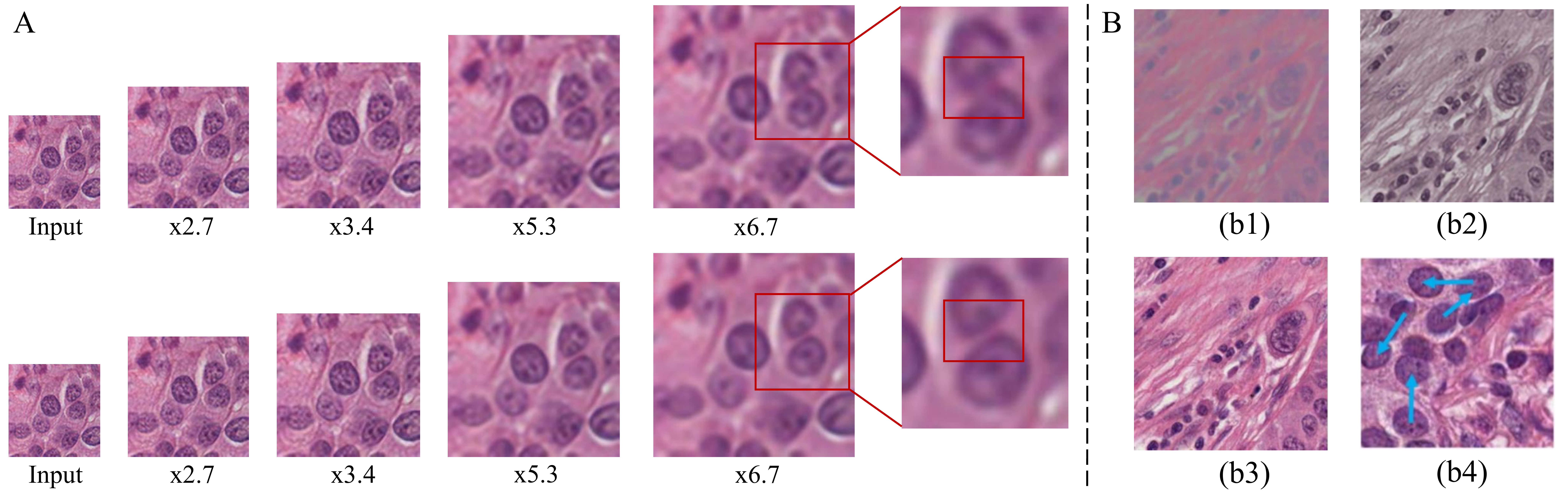}
    \caption{A. Non-integral magnification results of LIIF (upper row) and our ISTE (lower row). As shown in the red box, at the magnification of $\times$6.7, two cells are connected due to 
    blurring in the image generated by LIIF, while they are still separated in the image generated by ISTE. B. (b1). Pixels decoded by LPD (b2). Textures decoded by LTD (b3). Ground truth 
    image (b4). Visualization of texture similarity retrieval for the STF module, where the starting position of the blue arrow indicates the position of the retrieved texture 
    feature $F_{\mathrm{TL}}$ in the STF module, and the position pointed by the arrow is the position of the pixel feature that needs to be fused with the retrieved texture feature $F_{\mathrm{TL}}$ enhancement.
    }
    \label{figure4}
\end{figure*}

\begin{figure*}[htbp]
    \centering
    \includegraphics[scale=0.17]{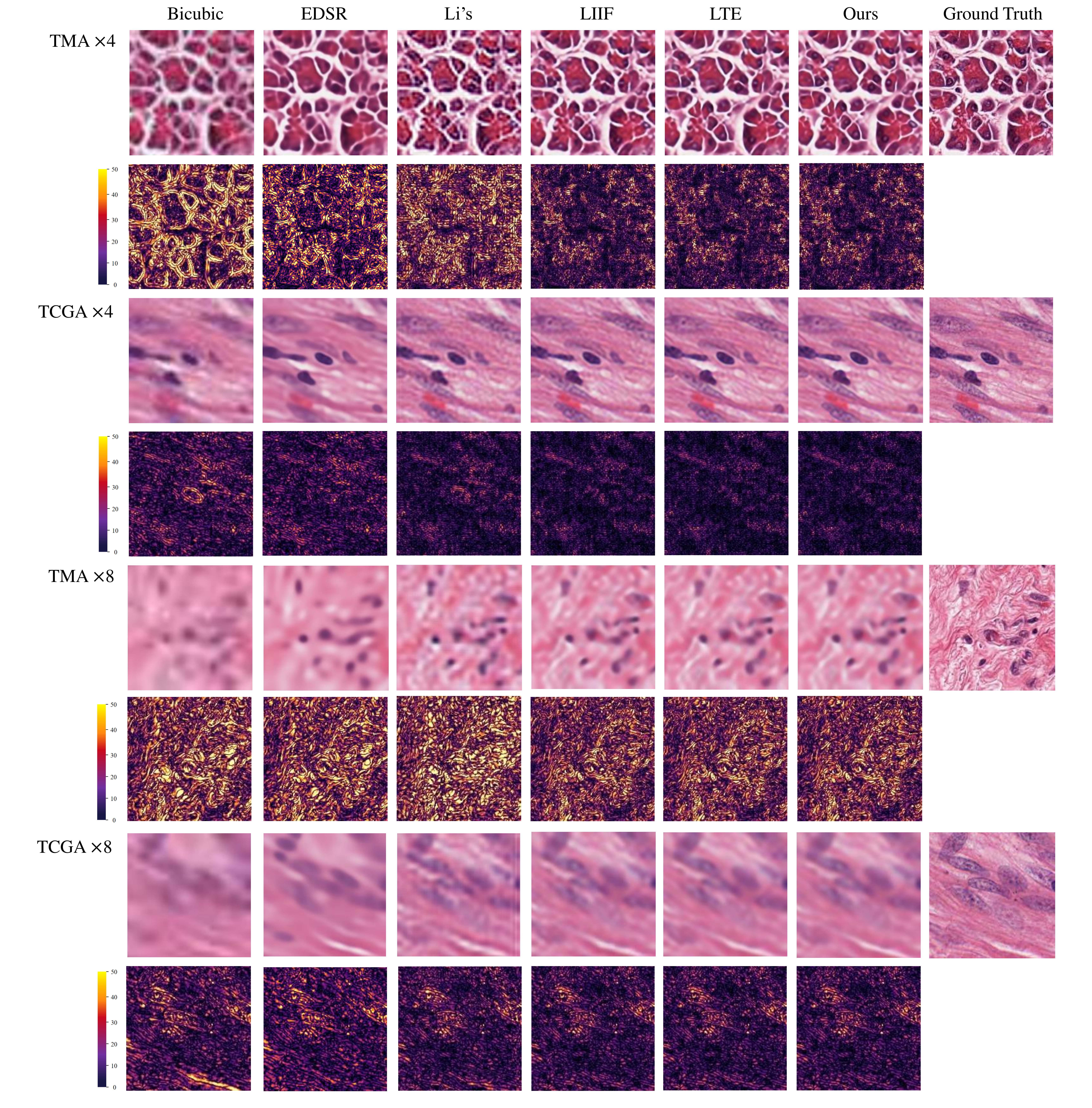}
    \caption{More visual comparisons with error maps of different methods on the TMA and TCGA datasets at 4$\times$ and 8$\times$ magnifications. The error maps represent the absolute 
        values of the error between the outputs and the ground truth images, with brighter colors representing larger errors.}
    \label{figure5}
\end{figure*}

\begin{figure*}[htbp]
    \centering
    \includegraphics[scale=0.17]{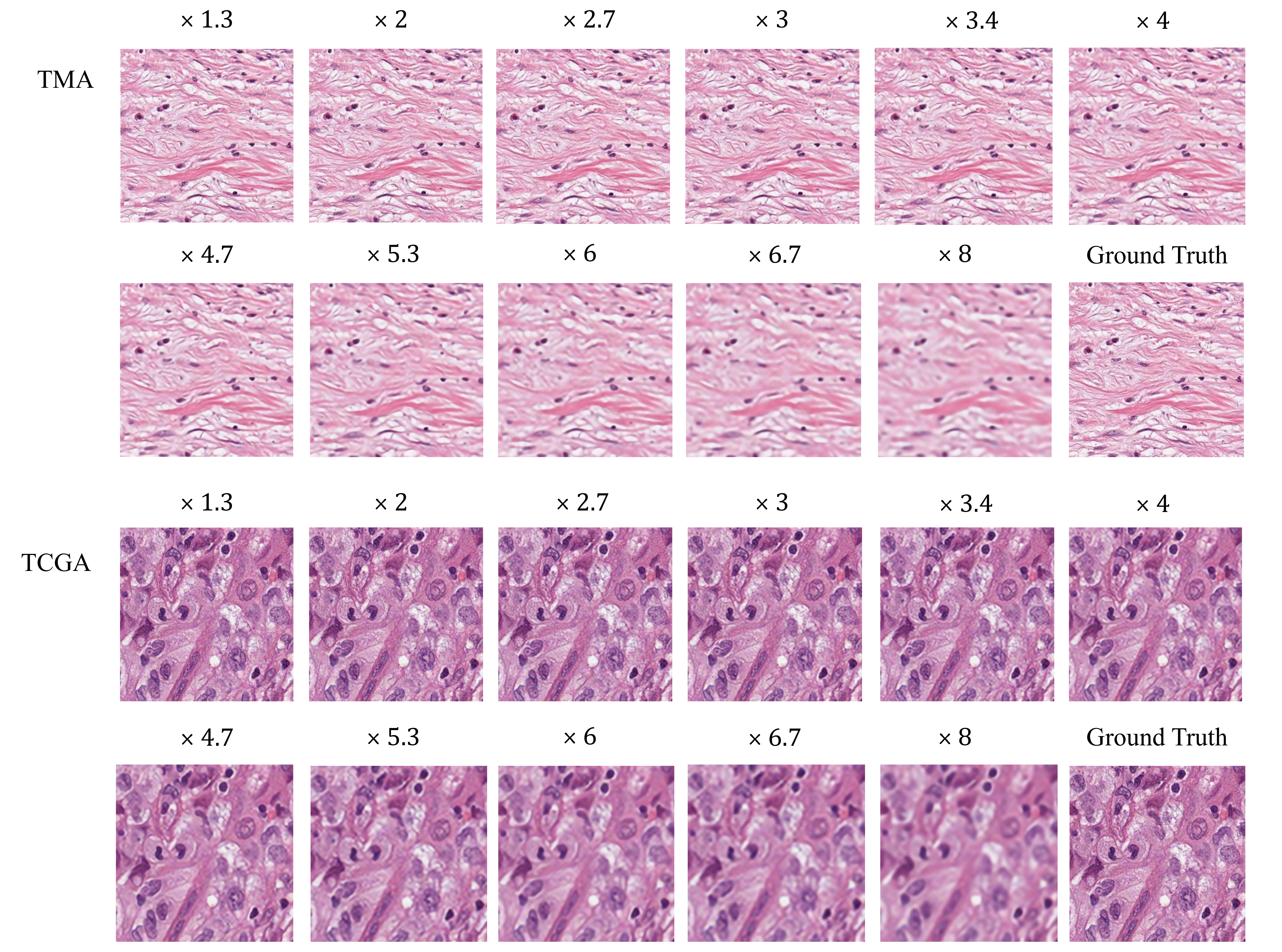}
    \caption{More visual results of our method at arbitary magnifications on the TMA and TCGA datasets.}
    \label{figure6}
\end{figure*}

\begin{figure*}[htbp]
    \centering
    \includegraphics[scale=0.17]{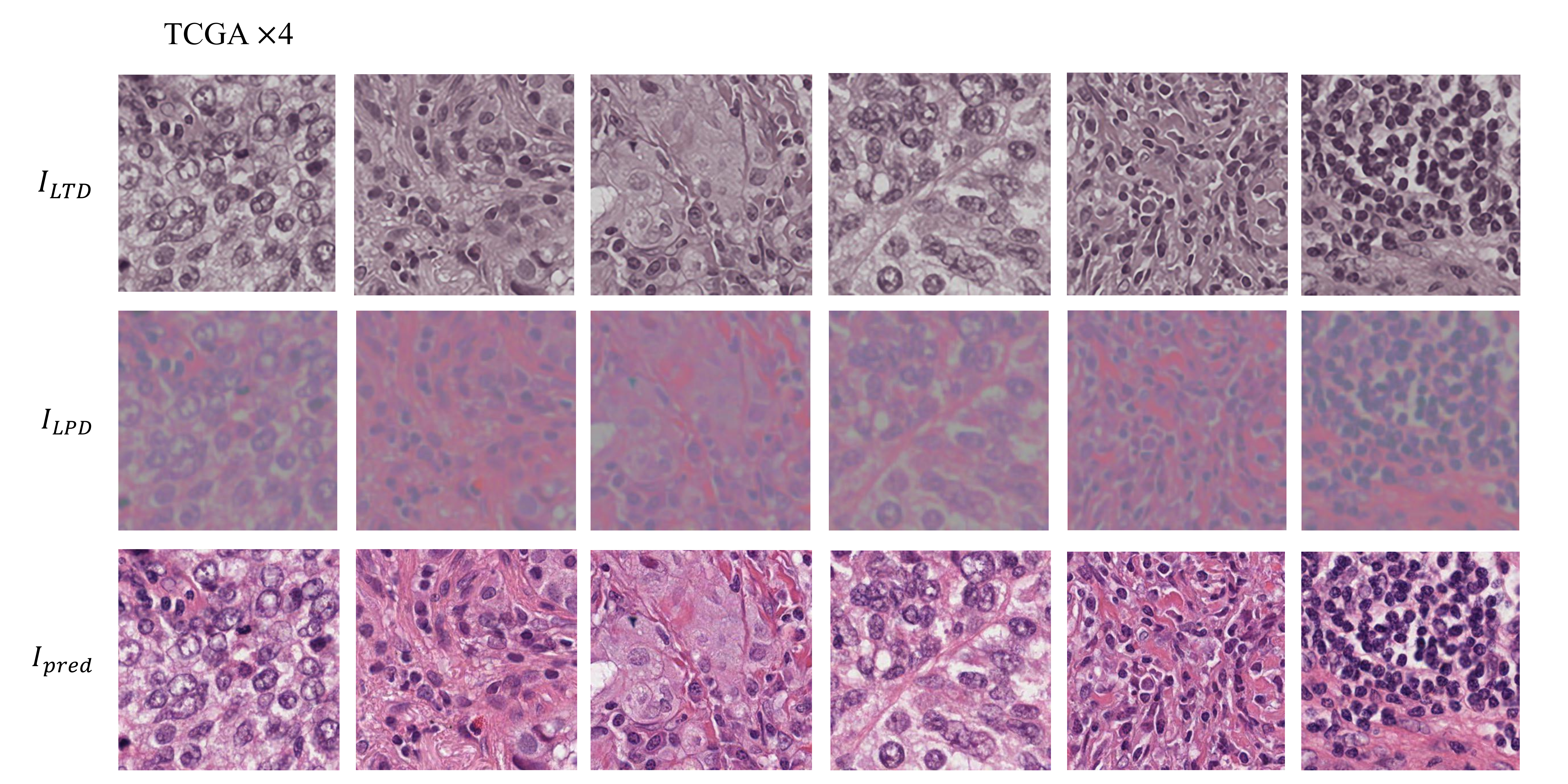}
    \caption{More visual results of the ablation study on the TCGA dataset. The first row: pixels decoded by LPD. The second row: textures decoded by LTD. The third row: final predictions.}
    \label{figure7}
\end{figure*}

We designed four variants of the network for ablation experiments on the TCGA dataset, and Table \ref{table2} shows the results. The main variants are as follows: \textbf{(2) w/o LFI}, remove the Local Feature Interactor module from our model; \textbf{(3) w/o STF}, remove the Self-Texture Fusion module from our model, 
and add the pixel information decoded by the Pixel Feature directly through the Local Pixel Decoder to the output of the Local-Texture Decoder module; \textbf{(4) w/o LTD}, remove the Local-Texture Decoder 
module from our model, and the features are directly decoded into pixel information as output by Local Pixel Decoder after the interaction between Pixel Feature and Texture Feature. The results 
demonstrate the efficiency of each component of ISTE. We also visualized the effect of texture transfer of STF. Fig. \ref{figure4} B (b4) indicates the texture transfer during one training iteration, where the
 blue arrows represent the direction of texture transfer. This demonstrates that our STF module does play a role in texture transfer. Further, we visualized the pixels decoded by LPD and the textures 
 decoded by LTD. As shown in Fig. \ref{figure4} B (b1-b3), the pixel information decoded with LPD alone is relatively smooth and lacks high-frequency texture details, while the texture information decoded by LTD 
 shows clear contours and textures of cells. This further illustrates the importance of spatial domain-based texture enhancement with LTD.
 More visual results of the ablation study on the TCGA dataset are shown in Fig. \ref{figure7}.

 \begin{table}[ht]
    \centering
    \caption{Results of ablation studies on the TCGA dataset.}
      \scalebox{0.60}{
        \setlength{\tabcolsep}{1.2mm}{
\begin{tabular}{
>{\columncolor[HTML]{FFFFFF}}c |
>{\columncolor[HTML]{FFFFFF}}c 
>{\columncolor[HTML]{FFFFFF}}c 
>{\columncolor[HTML]{FFFFFF}}c |
>{\columncolor[HTML]{FFFFFF}}c 
>{\columncolor[HTML]{FFFFFF}}c 
>{\columncolor[HTML]{FFFFFF}}c |
>{\columncolor[HTML]{FFFFFF}}c 
>{\columncolor[HTML]{FFFFFF}}c 
>{\columncolor[HTML]{FFFFFF}}c |
>{\columncolor[HTML]{FFFFFF}}c 
>{\columncolor[HTML]{FFFFFF}}c 
>{\columncolor[HTML]{FFFFFF}}c |
>{\columncolor[HTML]{FFFFFF}}c 
>{\columncolor[HTML]{FFFFFF}}c 
>{\columncolor[HTML]{FFFFFF}}c }
\toprule
\multicolumn{1}{l|}{\cellcolor[HTML]{FFFFFF}} & \multicolumn{3}{c|}{\cellcolor[HTML]{FFFFFF}×2}                                                              & \multicolumn{3}{c|}{\cellcolor[HTML]{FFFFFF}×3}                                       & \multicolumn{3}{c|}{\cellcolor[HTML]{FFFFFF}×4}                                                                        & \multicolumn{3}{c|}{\cellcolor[HTML]{FFFFFF}×6}                                                                        & \multicolumn{3}{c}{\cellcolor[HTML]{FFFFFF}×8}                                                                          \\ \cline{2-16} 
\multicolumn{1}{l|}{\cellcolor[HTML]{FFFFFF}} & PSNR↑                                 & FID↓                                 & SSIM↑                                  & PSNR↑                                 & FID↓ & SSIM↑                                  & PSNR↑                                 & FID↓                                  & SSIM↑                                  & PSNR↑                                 & FID↓                                  & SSIM↑                                  & PSNR↑                                 & FID↓                                   & SSIM↑                                  \\ \toprule
\textbf{Ours}                                 & {\color[HTML]{FF0000} \textbf{38.28}} & {\color[HTML]{FF0000} \textbf{1.22}} & {\color[HTML]{FF0000} \textbf{0.9748}} & {\color[HTML]{FF0000} \textbf{32.46}} & 2.92 & {\color[HTML]{FF0000} \textbf{0.9111}} & {\color[HTML]{FF0000} \textbf{29.33}} & 17.42                                 & {\color[HTML]{FF0000} \textbf{0.8287}} & {\color[HTML]{FF0000} \textbf{25.64}} & {\color[HTML]{FF0000} \textbf{88.00}} & 0.6617                                 & {\color[HTML]{FF0000} \textbf{23.74}} & 119.98                                 & 0.5578                                 \\ \toprule
w/o LFI                                       & 36.69                                 & 1.29                                 & 0.9727                                 & 31.90                                 & 3.06 & 0.9095                                 & 28.96                                 & 18.29                                 & 0.8250                                 & 25.50                                 & 89.36                                 & 0.6630                                 & 23.67                                 & 120.76                                 & 0.5591                                 \\
w/o  STF                                      & 36.82                                 & 1.25                                 & 0.9737                                 & 31.94                                 & 2.98 & 0.9103                                 & 29.05                                 & {\color[HTML]{FF0000} \textbf{17.22}} & 0.8274                                 & 25.57                                 & 88.47                                 & {\color[HTML]{FF0000} \textbf{0.6658}} & 23.74                                 & 121.80                                 & {\color[HTML]{FF0000} \textbf{0.5619}} \\
w/o  LTD                                      & 36.76                                 & 1.29                                 & 0.9737                                 & 31.90                                 & 3.04 & 0.9099                                 & 29.02                                 & 17.62                                 & 0.8264                                 & 25.52                                 & 87.54                                 & 0.6639                                 & 23.69                                 & {\color[HTML]{FF0000} \textbf{119.12}} & 0.5600                                 \\ \bottomrule
\end{tabular}}}
    \label{table2}
\end{table}

\section{Conclusion}
In this paper, we propose a novel dual-branch framework to achieve arbitrary-scale pathology image super-resolution for the first time. The proposed method outperforms SOTA natural and pathology image super-resolution methods in extensive experiments.

\bibliographystyle{splncs04}
\bibliography{ISTE}

\end{document}